\documentclass{article}
\usepackage[T1]{fontenc} 
\usepackage[utf8]{inputenc} 
\usepackage{ismir,amsmath,cite,url}
\usepackage{graphicx}
\usepackage{xcolor}
\usepackage{subfiles}
\usepackage{amsfonts}
\usepackage{float}
\usepackage{paralist}
\usepackage{units}
\usepackage{microtype}
\usepackage{caption, subcaption}
\usepackage{multirow}
\usepackage{boldline}

\usepackage{lineno}

\title{Audio Embeddings as Teachers for Music Classification}

\twoauthors
 {Yiwei Ding} {Music Informatics Group \\ Georgia Institute of Technology \\ {\tt yding402@gatech.edu}}
 {Alexander Lerch} {Music Informatics Group \\ Georgia Institute of Technology \\ {\tt alexander.lerch@gatech.edu}}

\def\authorname{Y. Ding and A. Lerch}

\begin{document}

\maketitle
\begin{abstract}
Music classification has been one of the most popular tasks in the field of music information retrieval.
With the development of deep learning models, the last decade has seen impressive improvements in a wide range of classification tasks.
However, the increasing model complexity makes both training and inference computationally expensive.
In this paper, we integrate the ideas of transfer learning and feature-based knowledge distillation and systematically investigate using pre-trained audio embeddings as teachers to guide the training of low-complexity student networks.
By regularizing the feature space of the student networks with the pre-trained embeddings, the knowledge in the teacher embeddings can be transferred to the students.
We use various pre-trained audio embeddings and test the effectiveness of the method on the tasks of musical instrument classification and music auto-tagging.
Results show that our method significantly improves the results in comparison to the identical model trained without the teacher's knowledge.
This technique can also be combined with classical knowledge distillation approaches to further improve the model's performance.

\end{abstract}

\section{Introduction}\label{sec:introduction}

    The classification of music has always been a widely popular task in the field of Music Information Retrieval (MIR). Music classification serves as an umbrella term for a variety of tasks, including music genre classification \cite{tzanetakis2002musical}, musical instrument classification \cite{humphrey2018openmic}, and music auto-tagging \cite{law2009evaluation}. 
    The last decade has seen dramatic improvements in a wide range of such music classification tasks due to the increasing use of artificial neural networks \cite{gururani2019attention, koutini2021receptive, choi2016automatic, kim2018sample}.
    
    One major contributing factor to these impressive accomplishments is the increased algorithmic complexity of the machine learning models which {also means that the training process requires an increased amount of data.}
    As not all tasks have this abundance of annotated data, transfer learning has been widely and successfully applied to various music classification tasks \cite{choi2017transfer}.
    In transfer learning, a model is first pre-trained on a large-scale dataset for a (source) task that is somewhat related to the (target) task and then fine-tuned with a comparably smaller dataset of the target task \cite{pan2010survey}.
    This enables knowledge to be transferred across datasets and tasks.
    Transfer learning has been repeatedly shown to result in state-of-the-art performance for a multitude of MIR tasks \cite{koutini2022efficient, kong2020panns, mccallum2022supervised}.

    Another side effect of the increasing model complexity is the slow inference speed.
    One way to address this issue is model compression by means of knowledge distillation. Here, a low-complexity (student) model is trained while leveraging the knowledge in the high-complexity (teacher) model \cite{hinton2015distilling, ba2014deep}.
    The teacher-student paradigm has met with considerable success in reducing the model complexity while minimizing performance decay \cite{yu2019learning, touvron2021training}.
    
    In this study, we integrate ideas and approaches from both transfer learning and knowledge distillation and apply them to the training of low-complexity networks to show the effectiveness of knowledge transfer for music classification tasks.
    More specifically, we utilize pre-trained audio embeddings as teachers to regularize the feature space of low-complexity student networks during the training process.
    Thus, the main contributions of this paper are a systematic study of
    \begin{compactitem}
        \item   the effectiveness of various audio embeddings as teachers for knowledge transfer,
        \item   different ways to apply the knowledge transfer from teachers to students, and
        \item   the impact of data availability on the performance of the investigated systems.
    \end{compactitem}
    The models and experiments are publicly available as open-source code.\footnote{\url{https://github.com/suncerock/EAsT-music-classification}. Last accessed on June 21, 2023.}

\section{Related Work}\label{sec:related_works}

    This section first briefly introduces transfer learning and knowledge distillation, which are both often used to transfer knowledge between tasks and models, respectively, and then surveys the application of feature space regularization in the training of neural networks.

\subsection{Transfer Learning}

    In transfer learning approaches, a model is pre-trained on a source task with a large dataset and subsequently fine-tuned on a (different but related) target task with a (typically smaller) dataset \cite{pan2010survey}.
    By utilizing the knowledge learned from the source task, models trained following the transfer learning paradigm can often achieve significantly better results than the same models trained directly on the target task \cite{kolesnikov2020big};
    this is especially the case if these models have a large number of parameters and the training data for the target task is limited.
    In the case where fine-tuning the whole model might be too computationally expensive, another way to do transfer learning is to use the pre-trained embeddings and train only the classification head. This allows for a separation of the tasks of computing the embeddings and the classification itself.

    Transfer learning has been successfully applied to a wide variety of areas ranging from computer vision \cite{deng2009imagenet, he2016deep} to natural language processing \cite{devlin2018bert}.
    In MIR, transfer learning has been used for a multitude of target tasks \cite{choi2017transfer, kong2020panns, alonso2020deep, koutini2022efficient}.
    Besides fine-tuning the whole model, pre-trained embeddings such as VGGish \cite{hershey2017cnn} and Jukebox \cite{dhariwal2020jukebox} have also shown good performance on many tasks including auto-tagging \cite{mccallum2022supervised, castellon2021codified}, instrument classification \cite{gururani2019attention, mccallum2022supervised}, and music emotion recognition \cite{koh2021comparison, bogdanov2022musav, mccallum2022supervised, castellon2021codified}.
    
    One disadvantage of transfer learning is the slow inference speed. In most cases, the model has a large number of parameters, which means that both fine-tuning (if done on the whole model) and inference potentially lead to a high computational workload.

\subsection{Knowledge Distillation}

    Approaches for knowledge distillation aim at model compression, i.e., reducing the complexity of the network. 
    The knowledge of a (usually high-complexity) pre-trained network (the teacher) is transferred to a different (low-complexity) network (the student) during the training phase, in which the student not only learns from the ground truth labels but also from the teacher predictions.
    This is achieved by adding a  ``distillation loss'' term to the student's loss function to learn from the teacher's prediction \cite{ba2014deep, hinton2015distilling}.
    
    The most popular distillation loss is the Kullback-Leibler divergence between the logits of the student and the teacher, with a hyperparameter called temperature to soften the probability distribution of the teacher's prediction over classes \cite{hinton2015distilling}.
    The soft target provides more ``dark'' knowledge than the ground truth hard label \cite{szegedy2016rethinking, muller2019does}.
    The Pearson correlation coefficient has also been proposed as a distance measure between the logits as an alternative to the Kullback-Leibler divergence \cite{huang2022knowledge}.

    Besides learning from logits, the student network can also try to learn from the feature map from the intermediate layers of the teacher network \cite{romero2015fitnets, heo2019comprehensive, peng2019correlation}.
    As the feature maps of the student and teacher do not necessarily share the same dimension and the same size, a variety of ways to match the feature space of the student and the teacher have been proposed \cite{heo2019comprehensive, yim2017gift, kim2018paraphrasing}.
    Therefore, feature-based knowledge distillation has more flexibility than the logits-based traditional approach, which, at the same time, also makes it more challenging to find the best way of matching the feature space \cite{gou2021knowledge, wang2021knowledge}.
    
\subsection{Feature Space Regularization}

    Feature-based knowledge distillation is a technique of regularizing the feature space of the network during training.
    Besides knowledge distillation, there exists a wide variety of other ways to implement regularization. One example is contrastive learning, which aims at contrasting the features of instances with positive labels against negative labels \cite{dosovitskiy2014discriminative, chen2020simple}.
    Contrastive learning has been shown to improve the performance of neural networks on music auto-tagging \cite{spijkervet2021contrastive, alonso2022music} and music performance assessment \cite{seshadri2021improving}.

    Regularizing the feature space using pre-trained audio embeddings has also been reported to be effective in music classification \cite{hung2022feature_classification} and music source separation \cite{hung2022feature_source}, where Hung and Lerch proposed to use pre-trained embeddings to help structure the latent space during training.
    This technique is similar to but different from both transfer learning and knowledge distillation.
    In transfer learning, the same model is used on two different datasets, and a typical setting is that knowledge from the large dataset will be transferred to the small dataset.
    In knowledge distillation, only one dataset is used and the typical setting is that the knowledge will be transferred from a large model to a small model.
    In comparison, regularizing the feature space using embeddings requires neither the dataset nor the model to be the same, yet still allows to transfer knowledge learned by the teacher model from a large dataset to the low-complexity student network for a different (small) dataset.

    \begin{figure*}[htbp]
     \centering
     \includegraphics[width=0.95\linewidth]{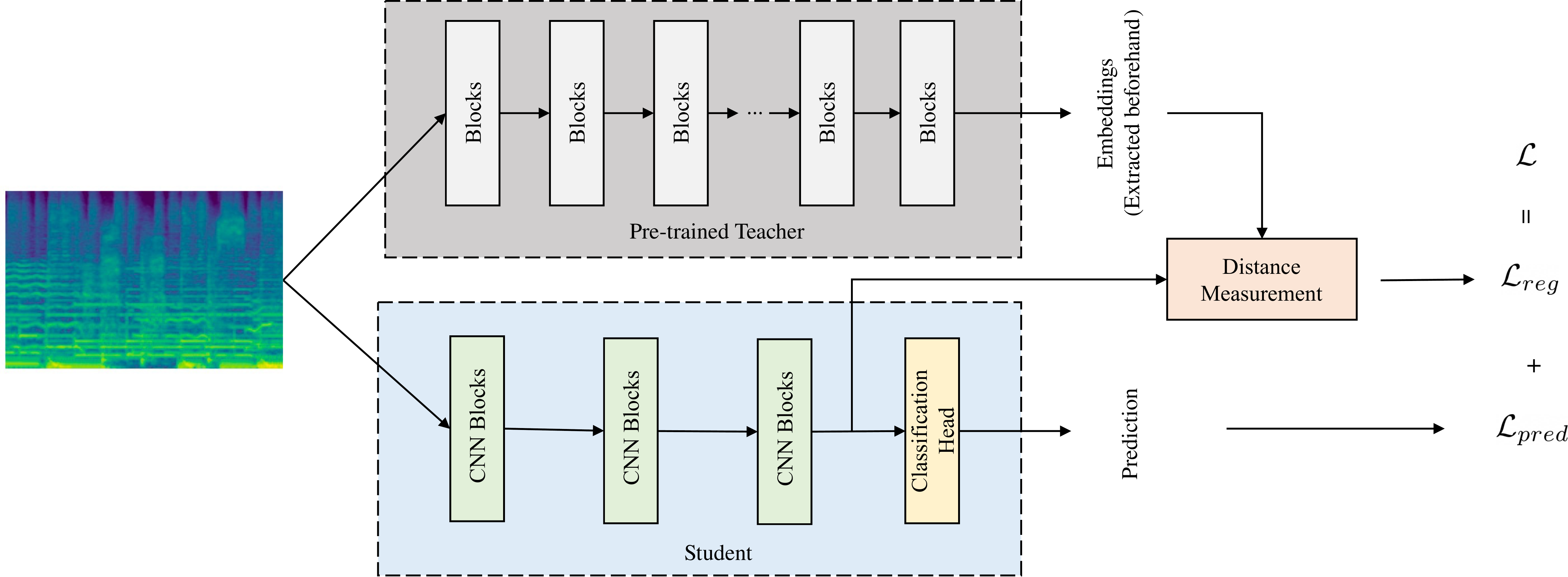}
     \caption{Overall pipeline of training a model by using pre-trained embeddings as teachers. The training loss is a weighted sum (weighting factor omitted in the figure) of prediction loss and regularization loss. The regularization loss measures the distance between pre-trained embedding and the output feature map after the feature alignment. During inference, only the bottom part with the blue background is used.}
     \label{fig:pipeline}
    \end{figure*}
    
\section{Methods}\label{sec:proposed_methods}

    Inspired by the promising preliminary results of prior work \cite{hung2022feature_classification}, we integrate the idea of transfer learning and knowledge distillation by using pre-trained audio embeddings as teachers to regularize the feature space of the student network during training.
    The overall pipeline is illustrated in \figref{fig:pipeline}.
    
\subsection{Loss Function}
    Similar to knowledge distillation \cite{hinton2015distilling}, we rewrite our loss function as 
    \begin{equation}
        \mathcal{L} = (1 - \lambda)\mathcal{L}_\mathrm{pred} + \lambda\mathcal{L}_\mathrm{reg}
    \end{equation}
    where $\mathcal{L}_\mathrm{pred}$ is the loss function for conventional neural network training, $\mathcal{L}_\mathrm{reg}$ is the loss function that measures the distance between the student network's feature map and the pre-trained embeddings, and $\lambda\in[0, 1]$ is a weighting hyper-parameter.

\subsection{Regularization Location}
    Different stages in a neural network output different feature maps, and the optimal location to apply regularization continues to be controversially discussed in feature-based knowledge distillation \cite{wang2021knowledge}.
    In this study, we investigate either regularizing only the final feature map before the classification head as shown in \figref{fig:pipeline} or regularizing the feature maps at all stages of the student network. 

\subsection{Feature Alignment}
    To measure the distance between the {student feature map $l\in\mathbb{R}^{T_\mathrm{s} \times C_\mathrm{s}}$ and the pre-trained teacher embeddings $v\in\mathbb{R}^{T_\mathrm{t} \times C_\mathrm{t}}$ which might have different numbers of time frames (i.e., $T_\mathrm{s} \neq T_\mathrm{t}$)}, we first align the intermediate feature map with the pre-trained embeddings in time by repeating the one with fewer time frames, then compute the distance for each frame and finally average them along the time axis. 

\subsection{Distance Measure}\label{subsec:distance_measure}
    Considering that pre-trained embeddings and feature maps have often different dimensionalities, the use of distance measures that are independent of dimensionality allows for easier application.

    \subsubsection{Cosine Distance Difference}
    Cosine distance difference\footnote{has been referred to in previous work as Distance-based Regularization (Dis-Reg)\cite{hung2022feature_classification, hung2022feature_source}.} as proposed in previous work \cite{hung2022feature_classification, hung2022feature_source} measures the difference in the cosine distance between pairs of samples. Given $n$ pairs of samples of single-time-frame features $l_1, l_2, ..., l_n$ and pre-trained embeddings $v_1, v_2, ..., v_n$, the cosine distance difference for one pair is
    \begin{equation}
        D_{ij} = |d_\mathrm{cos}(l_i, l_j) - d_\mathrm{cos}(v_i, v_j)|,
    \end{equation}
    and the distance for this time frame is averaged among all pairs.

    \subsubsection{Distance Correlation}
    Distance correlation was proposed as a generalization of classical correlation to measure the independence between two random vectors in arbitrary dimensions \cite{szekely2007measuring}.
    It is capable of handling features of different dimensionality; furthermore, correlation-based distance measures have been shown to be effective in knowledge distillation \cite{peng2019correlation, huang2022knowledge}.
    Using the same notation as above, we define
    \begin{align}
        a_{ij} &= \Vert l_i - l_j \Vert, \label{eq1} \\
        \bar{a}_{i.} &= \dfrac{1}{n}\sum_{j=1}^n a_{ij}, \quad
        \bar{a}_{.j} = \dfrac{1}{n}\sum_{i=1}^n a_{ij}, \quad
        \bar{a}_{..} = \dfrac{1}{n^2}\sum_{i,j=1}a_{ij}\\
        A_{ij} &= a_{ij} - \bar{a}_{i.} - \bar{a}_{.j} + \bar{a}_{..}
    \end{align}
    where $i, j \in \{1, 2, ..., n\}$, and similarly, $b_{ij} = \Vert v_i - v_j \Vert$ and $B_{ij} = b_{ij} - \bar{b}_{i.} - \bar{b}_{.j} + \bar{b}_{..}$.\footnote{Eq.~(\ref{eq1}) uses 2-norm following the implementation in \url{https://github.com/zhenxingjian/Partial_Distance_Correlation}.}
    The distance for the time frame is then 
    \begin{equation}
        \mathcal{L}_\mathrm{reg} = 1 - \mathcal{R}_n^2(l, v) = 1 - \dfrac{\mathcal{V}^2_n(l, v)}{\sqrt{\mathcal{V}^2_n(l, l)\mathcal{V}^2_n(v, v)}}
    \end{equation}
    where 
		\begin{eqnarray}
        &\mathcal{V}^2_n(l, l) = \dfrac{1}{n^2}\sum\limits_{i,j=1}^nA_{ij}^2,\quad \mathcal{V}^2_n(v, v) = \dfrac{1}{n^2}\sum\limits_{i,j=1}^nB_{ij}^2, &\nonumber\\
        &\mathcal{V}^2_n(l, v) = \dfrac{1}{n^2}\sum\limits_{i,j=1}^nA_{ij}B_{ij}. &\nonumber
		\end{eqnarray}

    Note that $\mathcal{V}^2_n(l, l)$ and $\mathcal{V}^2_n(v, v)$ will be $0$ if and only if all the {$n$ samples of features (or embeddings) within one batch} are identical \cite{szekely2007measuring}, which we assume not to occur here. 

    To optimize both distance measures during training, block stochastic gradient iteration is used, which means that the distance is computed over mini-batches instead of the whole dataset \cite{xu2015block, zhen2022versatile}.
    With stochastic approximation, the computational complexity of the distance measure for $n$ samples is reduced from $\mathcal{O}(n^2)$ to $\mathcal{O}(mn)$ where $m$ is the batch size.

    It is worth mentioning that both distance measures ensure that if the distance is zero, the feature maps would differ from the pre-trained embeddings by only an orthogonal linear transformation, which can be modeled in a single linear layer.
    Therefore, if the regularization loss is zero, the student would have the same performance as the teacher in classification.

\section{Experimental Setup}\label{sec:experiments}

    We test the effectiveness of using pre-trained embeddings as teachers on two different tasks, datasets, and models with four different pre-trained embeddings as follows.
    
\subsection{Tasks, Datasets, Models, and Metrics}

\begin{table*}[!h]
    \begin{center}
    \begin{tabular*}{\textwidth}{l|@{\extracolsep{\fill}}cccccccccc}
        \hline
        \hline
        
        
        \multirow{2}{*}{\textbf{OpenMIC}}
                & \multicolumn{2}{c}{None}
                & \multicolumn{2}{c}{VGGish}   & \multicolumn{2}{c}{OpenL3}
                & \multicolumn{2}{c}{PaSST}    & \multicolumn{2}{c}{PANNs}\\
        \cline{2-11}
                & mAP           & F1
                & mAP           & F1            & mAP           & F1
                & mAP           & F1            & mAP           & F1\\
        \hline
        CP ResNet* \cite{koutini2021receptive}  & .819          & .809
                & -     & -     & -     & -     & -     & -     & -     & -\\
        SS CP ResNet* \cite{koutini2021receptive}  & .831          & .822
                & -     & -     & -     & -     & -     & -     & -     & -\\
        \hline
        Teacher\textsubscript{LR} & - & -
                & .803          & .799          & .803          & .798
                & \textbf{.858} & \textbf{.837} & .853          &\textbf{.834}\\
        KD (w/ mask) ** & - & -
                & .829          & .820          & .823          & .813
                & .851          & \underline{.834}          & .848          & .823\\
        \hline
        EAsT\textsubscript{Cos-Diff} & - & -
                & .838          & .824          & \underline{\textbf{.838}} & .820
                & .837          & .822          & .836          & .814\\
        EAsT\textsubscript{Final} & - & -
                & \underline{\textbf{.842}}     & \underline{\textbf{.828}} 
                & .835          & \underline{\textbf{.822}}
                & .847          & .830          & .849          & .828\\
        EAsT\textsubscript{All} & - & -
                & .836          & .823          & .835          & \underline{\textbf{.822}}
                & .845          & .827          & .845          & .827\\
        EAsT\textsubscript{KD} & - & -
                & .836          & .825          & .836          & .821
                & \underline{.852}              & \underline{.834}       
                & \underline{\textbf{.857}}     & \underline{.831}\\
        \hline
        \hline

        \\

        \hline
        \hline
        \multirow{2}{*}{\textbf{MagnaTagATune}}
                & \multicolumn{2}{c}{None}
                & \multicolumn{2}{c}{VGGish}   & \multicolumn{2}{c}{OpenL3}
                & \multicolumn{2}{c}{PaSST}    & \multicolumn{2}{c}{PANNs}\\
        \cline{2-11}
                & mAP           & AUC
                & mAP           & AUC            & mAP           & AUC
                & mAP           & AUC            & mAP           & AUC\\
        \hline
        FCN$^\dagger$    \cite{choi2016automatic}   & .429          & .900
                & -     & -     & -     & -     & -     & -     & -     & -\\
        Mobile FCN 
                & .437          & .905
                & -     & -     & -     & -     & -     & -     & -     & -\\
        \hline
        Teacher\textsubscript{LR} & -             & -
                & .433          & .903          & .403          & .890
                & \textbf{.473} & \textbf{.917} & \textbf{.460} & .911\\
        KD      & -             & -
                & .447          & .911          & .439          & .907
                & .454          & .912          & .448          & .909\\
        \hline
        EAsT\textsubscript{Cos-Diff} & -             & -
                & .446          & .906          & .438          & .907
                & .453          & .912          & .453          & .911\\
        EAsT\textsubscript{Final}
                & -             & -
                & .454          & \underline{\textbf{.912}} & .447          & .910
                & .459          & .912          & .449          & .909\\
        EAsT\textsubscript{All}
                & -             & -
                & \underline{\textbf{.455}}     & .911          
                & \underline{\textbf{.452}}     & \underline{\textbf{.911}}
                & .458          & .913          & .457          & .911\\
        EAsT\textsubscript{KD}
                & -             & -
                & .441          & .908          & .437          & .904
                & \underline{.461}              & \underline{.915}          
                & \underline{.459}              & \underline{\textbf{.912}}\\
        \hline
        \hline
    \end{tabular*}
    \end{center}
    \caption{Results on OpenMIC (above) and MagnaTagATune (below) dataset for different models regularized with different pre-trained embeddings. Best performances are in bold, and best results excluding the teachers are underlined. *Reported results \cite{koutini2021receptive}, SS means being trained with shake-shake regularization \cite{gastaldi2017shake}. **When using KD, the missing labels in OpenMIC were masked to avoid potentially adding more training data. $^\dagger$Results from the open-source re-implementation \cite{won2020eval}.}
    \label{tab:results}
\end{table*}

    \subsubsection{Musical Instrument Classification with OpenMIC}
    Musical instrument classification is a multi-label classification problem. We use the OpenMIC dataset \cite{humphrey2018openmic}, which provides weakly labeled audio snippets of length \unit[10]{s}.
    Following prior work \cite{gururani2019attention, chen2023music}, we use the suggested test set and randomly split 15\% of the training data as the validation set, resulting in 12,692 training observations, 2,223 validation observations, and 5085 test observations.
    To ensure a consistent sample rate, the audio is resampled to \unit[32]{kHz} \cite{koutini2021receptive, chen2023music}.
    As the dataset is not completely labeled, i.e., parts of the labels are missing and not labeled as positive or negative, the missing labels are masked out when computing the loss function as suggested in previous work \cite{chen2023music, koutini2021receptive, koutini2022efficient}.

    We use receptive field regularized ResNet (CP-ResNet) \cite{koutini2021receptive} for this task, as it reaches state-of-the-art performance when trained only on the OpenMIC dataset (i.e., neither trained with transfer learning nor trained with any knowledge distillation).
    CP-ResNet has a ResNet-like structure \cite{he2016deep} with an added hyper-parameter $\rho$ to control the maximum receptive field of the ResNet.
    {We set $\rho=7$ to match the setting which provides the best results in the original work \cite{koutini2021receptive}}.

    The results are reported with the metrics mean Average Precision (mAP) and F1-score.
    The F1-score is calculated in a macro fashion, which means that for each instrument, the F1-score is computed for both the positive labels and the negative labels and then averaged, and the final F1-score is the mean of the F1-scores of all instruments.
    The detection threshold for the prediction is set to 0.4 following previous work \cite{koutini2021receptive}.

    \subsubsection{Music Auto-Tagging with MagnaTagATune}
    Similar to musical instrument classification, music auto-tagging is also a multi-label classification problem.
    We use the MagnaTagATune dataset \cite{law2009evaluation} for this task, which comes with audio clips of approximately \unit[29.1]{s}. 
    Following previous work, we use only the top 50 labels and exclude all the songs without any positive label from the dataset \cite{won2019toward, kim2018sample}.
    For comparability, the data split is adopted from previous work, with audio files in the directories '0' to 'b' being the training set, 'c' being the validation set, and 'e' and 'f' being the test set \cite{won2020data, won2020eval}, resulting in 15,247 training clips, 1,529 validation clips, and 4,332 test clips. 

    We apply a modified fully convolutional neural network (FCN) \cite{choi2016automatic} to this task.
    It is the simplest model among the benchmark models for the MagnaTagATune dataset \cite{won2020eval} and consists of several convolution and max-pooling layers.
    To further reduce the complexity of the model, we apply the MobileNet-like modification \cite{howard2017mobilenets} to the network by breaking the $3\times3$ convolutions into depth-wise separable convolutions and $1\times1$ convolutions.

    The results are evaluated with mAP and ROC-AUC.

\subsection{Pre-trained Embeddings}
    \subsubsection{VGGish}
    VGGish \cite{hershey2017cnn} is a widely used embedding in MIR, with a VGG network \cite{simonyan2014very} being trained on a large number of Youtube videos.
    The open-source PyTorch implementation is used to extract VGGish features\footnote{\url{https://github.com/harritaylor/torchvggish}. Last accessed on April 4, 2023.} which by default extracts 128 principle components and then quantizes them to \unit[8]{bit}.
    The time resolution is \unit[960]{ms}.

    \subsubsection{OpenL3}
    The OpenL3 embedding \cite{cramer2019look, arandjelovic2017look} is trained on a music subset of AudioSet \cite{gemmeke2017audio} {in a self-supervised paradigm}.
    The audio embeddings are extracted using the open-source Python package \texttt{OpenL3}\footnote{\url{https://github.com/marl/openl3/tree/main}. Last accessed on April 4, 2023} with the dimensionality being 512.
    To keep consistent with VGGish, the time resolution is set to \unit[960]{ms}.

    \subsubsection{PaSST}
    PaSST \cite{koutini2022efficient} is a 7-layer transformer trained on AudioSet for acoustic event detection.
    It applies the structure of a vision transformer \cite{dosovitskiy2021an, touvron2021training} and proposes the technique of Patchout to make the training efficient.
    We use the open-source code\footnote{\url{https://github.com/kkoutini/PaSST/tree/main}. Last accessed on April 4, 2023.} released by the authors to extract the 768-dimensional embeddings.
    The time resolution is also set to \unit[960]{ms}.

    \subsubsection{PANNs}
    PANNs \cite{kong2020panns} include several convolutional neural networks and are also trained on AudioSet for acoustic event detection.
    We use the default CNN14 model from the official repository\footnote{\url{https://github.com/qiuqiangkong/audioset_tagging_cnn}. Last accessed on April 4, 2023. }. 
    The embedding dimensionality is 2048.
    Different from other embeddings, PANNs provide only one global embedding for each clip of audio.
    Pilot experiments have shown that extracting the embeddings for short segments and concatenating them does not improve performance.

\subsection{Systems Overview}
    The following systems are evaluated for comparison:
    \begin{compactitem}
    \setdefaultleftmargin{1em}{1em}{1em}{1em}{1em}{1em}
        \item Baseline: CP ResNet (on OpenMIC) and Mobile FCN (on MagnaTagATune) trained without any extra regularization loss.
        \item Teacher\textsubscript{LR}: logistic regression on the pre-trained embeddings (averaged along the time axis), which can be seen as one way to do transfer learning by freezing the whole model except for the classification head.
        \item KD: classical knowledge distillation where the soft targets are generated by the logistic regression.
        \item EAsT\textsubscript{Cos-Diff}  (for Embeddings-As-Teachers): feature space regularization as proposed by Hung and Lerch that uses cosine distance difference and regularizes only the final feature map  \cite{hung2022feature_classification}.
        \item EAsT\textsubscript{Final} and EAsT\textsubscript{All}: proposed systems based on distance correlation as the distance measure, either regularizing only at the final stage or at all stages, respectively.
        \item EAsT\textsubscript{KD}: a combination of classical knowledge distillation and our method of using embeddings to regularize the feature space. The feature space regularization is done only at the final stage.
    \end{compactitem}
    We perform a search of $\lambda$ for each of the EasT systems and choose the best-performing value on the validation set.\footnote{For all the hyperparameters, please refer to the config files in our GitHub.}

\section{Results}\label{sec:results}

    This section presents the results of different systems and their performance in the case of limited training data.

\subsection{Results on OpenMIC and MagnaTagATune}
    Table~\ref{tab:results} shows the results on the OpenMIC and the MagnaTagATune datasets.

    We can observe that the models trained with the extra regularization loss consistently outperform the non-regularized ones on both datasets, with all features, and all regularization methods. This means that the knowledge in the embeddings is successfully transferred to the student networks and consistently enhances the performance.
    
    Although EAsT\textsubscript{Final} appears to give better results on the OpenMIC dataset while EAsT\textsubscript{All} seems to have slightly better performance on the MagnaTagATune dataset, the difference between them is very small, meaning that the model does not benefit significantly from being regularized by pre-trained embeddings at earlier stages where the feature maps are still relatively low-level.

    The results for the teacher systems show that the older VGGish and OpenL3 embeddings are clearly outperformed by the more recently proposed embeddings PaSST and PANNs.
    In fact, the teacher systems for the newer embeddings perform so strongly that the students can rarely outperform them, while the student systems trained with VGGish and OpenL3 provide better results than the corresponding teachers.
    We can see that whether the teachers themselves have an excellent performance or not, students benefit from learning the additional knowledge from these embeddings, and the students' upper limit is not bounded by the performance of teachers.

    Comparing KD and the EAsT\textsubscript{Final} or EAsT\textsubscript{All} systems, we can see that with VGGish and OpenL3 embeddings, regularizing the feature space provides better results than simply using the teachers' soft targets.
    On the other hand, for the PaSST and PANNs embeddings, classical knowledge distillation provides competitive results.
    The possible reason is that the soft targets given by ``weak'' teachers might have provided too much incorrect information to the students while the high-quality soft targets generated by the ``strong'' teachers provide good guidance for the students' training.

    The combination system EAsT\textsubscript{KD} gives us better results with PaSST and PANNs embeddings (with the exception of no noteworthy improvement with the PaSST embedding on the OpenMIC dataset) while for VGGish and OpenL3 embeddings, the performance is not as good as EAsT\textsubscript{Final} or EAsT\textsubscript{All} in most cases.
    This observation is in accordance with our speculation that traditional knowledge distillation performs best with a ``strong'' teacher.
    While learning from audio embeddings benefits a student network even more in the presence of a ``strong'' teacher, learning from ``weak'' embeddings can still improve the model's performance.

\subsection{Comparison of Model Complexity}

    Table \ref{tab:results_complexity} lists the number of parameters as well as rough inference speed measurements\footnote{reference GPU: NVIDIA 2070 Super} of the models.

\begin{table}
    \begin{center}
    \begin{tabular*}{\columnwidth}{l|@{\extracolsep{\fill}}cc}
    \hline
    \hline
    Model       & Parameters (M)    & Iteration / s\\
    \hline
    VGGish      & 72.1              & 172.2\\
    OpenL3      & 4.69              & 117.9\\
    PaSST       & 86.1              & 18.7\\
    PANNs       & 79.7              & 70.6\\
    \hline
    Mobile FCN  & 0.34              & 319.3\\
    CP ResNet   & 5.52              & 205.3\\
    \hline
    \hline
    
    \end{tabular*}
    \end{center}
    \caption{Comparison of the model complexity.}
    \label{tab:results_complexity}

\end{table}

    The numbers of parameters only take the backbone structure (i.e., excluding the final classification head) into account so that it does not vary across datasets with different numbers of classes. Iterations per second are tested with 128$\times$1000 input spectrograms.

    We can see that Mobile FCN and CP ResNet are much faster in inference than pre-trained models.

\subsection{Limited Training Data}
\begin{figure}[htbp]
         \begin{subfigure}[b]{0.49\linewidth}
             \centering
             \includegraphics[width=\linewidth]{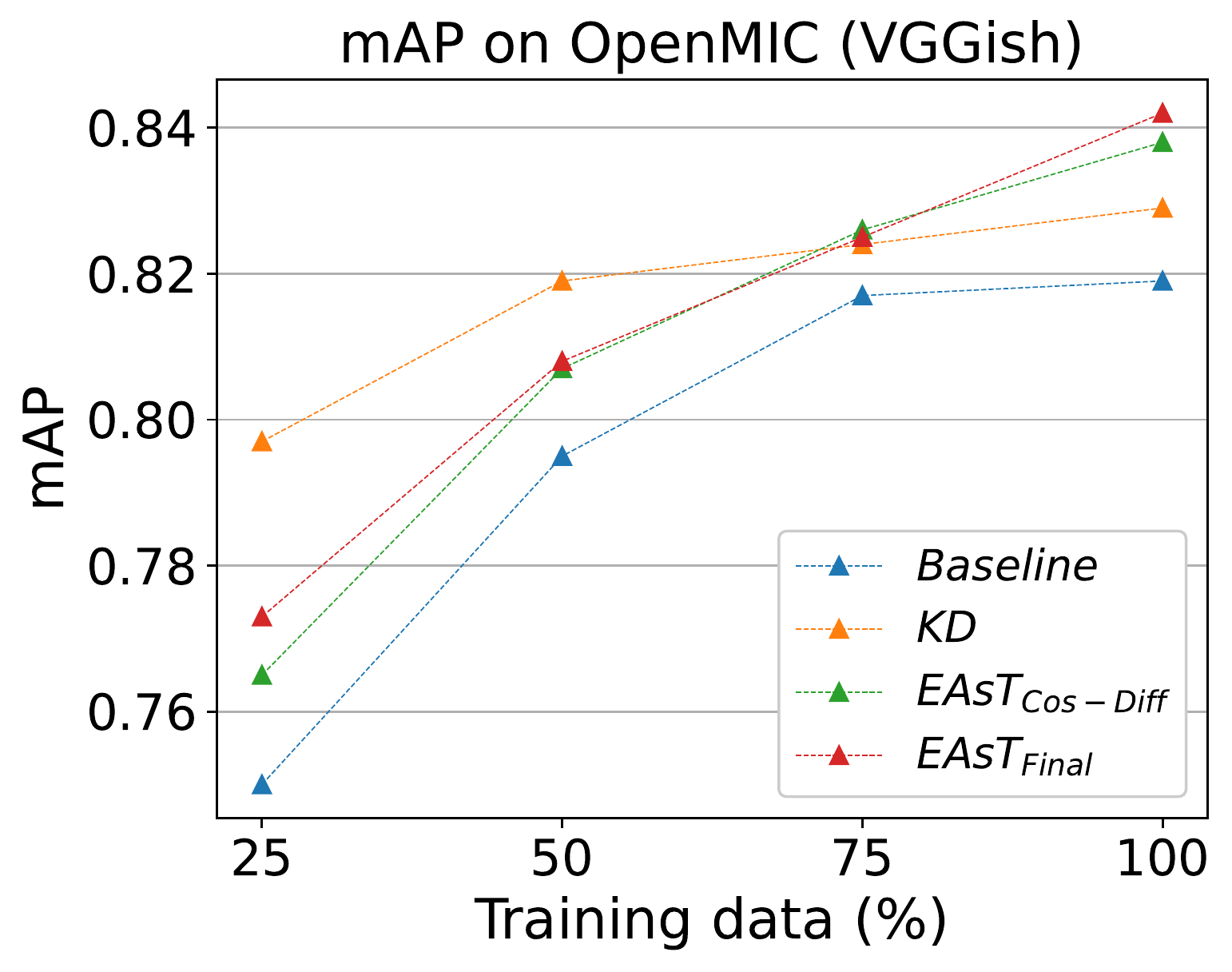}
             \label{fig:limited_training_data_openmic_vggish}
         \end{subfigure}
         \hfill
         \begin{subfigure}[b]{0.49\linewidth}
             \centering
             \includegraphics[width=\linewidth]{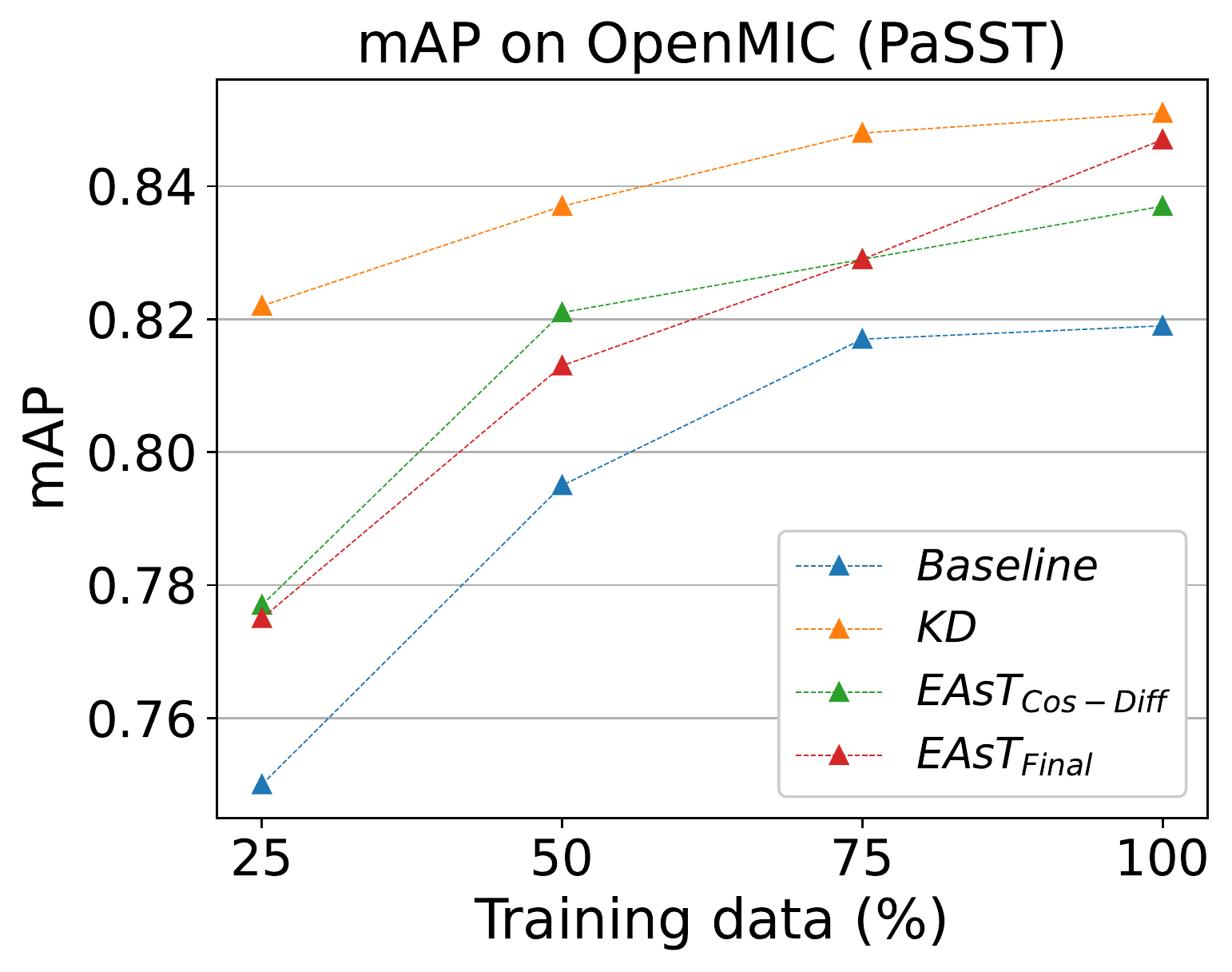}
             \label{fig:limited_training_data_openmic_passt}
         \end{subfigure}
         \hfill
         \begin{subfigure}[b]{0.49\linewidth}
             \centering
             \includegraphics[width=\linewidth]{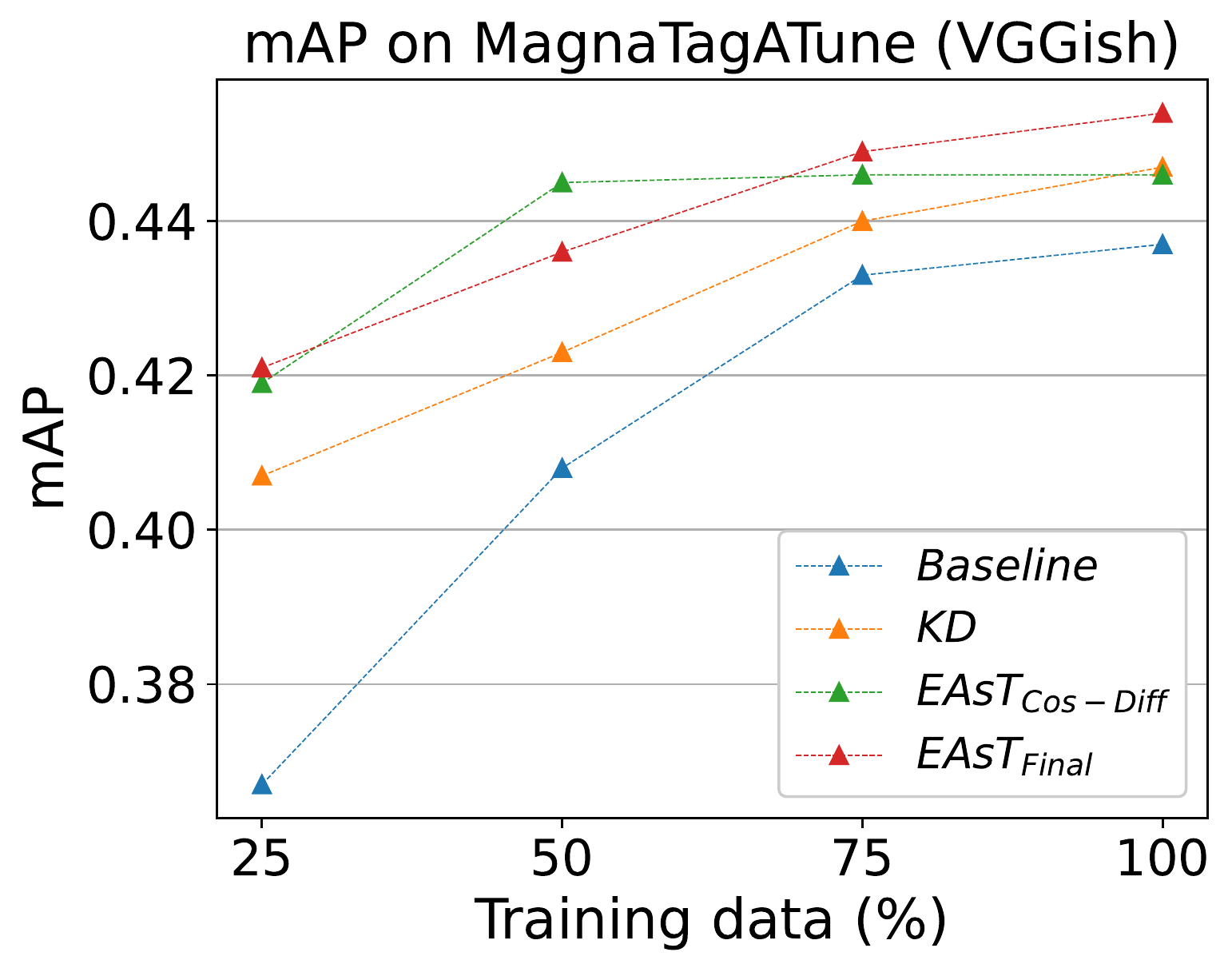}
             \label{fig:limited_training_data_mtat_vggish}
         \end{subfigure}
         \hfill
         \begin{subfigure}[b]{0.49\linewidth}
             \centering
             \includegraphics[width=\linewidth]{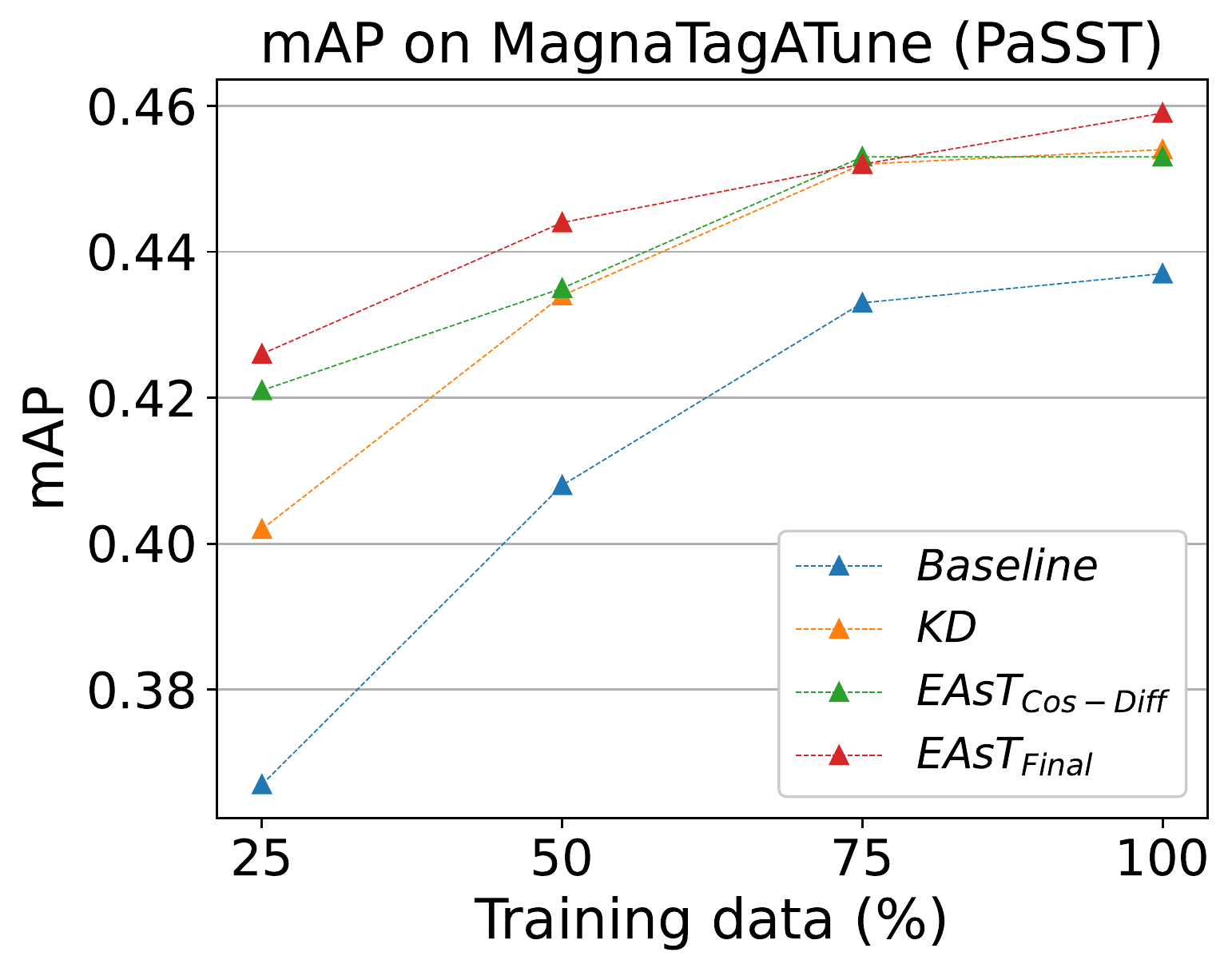}
             \label{fig:limited_training_data_mtat_passt}
         \end{subfigure}
     \caption{Results with limited training data on two datasets.}
     \label{fig:limited_training_data}
    \vspace{-1mm}
    \end{figure}

    To investigate the impact of limited training data on our methods, we present the system performances for reduced training data, i.e., for 25\%, 50\%, and 75\% of the original training data. The results are shown in \figref{fig:limited_training_data}.
    We use VGGish and PaSST as the pre-trained embeddings.
    
    We can observe that limiting the training data has the greatest impact on the baseline systems, which show the biggest performance drop.
    
    On the OpenMIC dataset, EAsT\textsubscript{Cos-Diff} and EAsT\textsubscript{Final} have similar decreases in mAP, and the KD system is less affected.
    An interesting finding is that when the VGGish embedding is used, KD shows better performance for limited data amounts while it is outperformed by EAsT\textsubscript{Cos-Diff} and EAsT\textsubscript{Final} when the whole OpenMIC dataset is available. This means using embeddings as teachers might still require a sufficient amount of data to have good guidance on the student models.

    On the MagnaTagATune dataset, however, the EAsT\textsubscript{Cos-Diff} and EAsT\textsubscript{Final} systems show less performance decay than either KD or the baseline when the training data is limited. This suggests that in our training settings, there is no certain answer to which method is least affected by the lack of training data, and the answer might be dependent on specific tasks, models, and data.




\section{Conclusion and Future Work}\label{sec:conclusion}


In this paper, we explored the use of audio embeddings as teachers to regularize the feature space of low-complexity student networks during training.
We investigated several different ways of implementing the regularization and tested its effectiveness on the OpenMIC and MagnaTagATune datasets.
Results show that using embeddings as teachers enhances the performance of the low-complexity student models, and the results can be further improved by combining our method with a traditional knowledge distillation approach.

Future work will investigate the performance of our method on a wider variety of downstream tasks {and embeddings}.
Moreover, as there have been a wide variety of models to extract audio and music embeddings, we speculate that using an ensemble of different pre-trained embeddings also has considerable potential.
Finally, the flexibility of feature-based knowledge distillation offers a wide range of possible algorithmic modifications. Our focus will be on evaluating different distance measures and regularizing the network using features from different stages of the teacher network instead of using only the output embeddings.

\bibliography{ISMIRtemplate}

%
%
  %
%
%

\end{document}